\documentclass[12pt]{article}
\usepackage{graphicx}

\newcommand {\be}{\begin{equation}} 
\newcommand{\ee}{\end{equation}}    

\def\dds1{\frac{\partial}{\partial s_1}}

\def\d{d\kern-0.8 ex\vrule height 1.3 ex depth-1.24 ex width 0.7 ex
\kern 0.15 ex}
\def\D{D\kern-1.7 ex\vrule height .87 ex depth-0.8 ex width 0.7 ex
\kern 0.95 ex}

\begin{document}
\baselineskip 20 pt

\begin{center}

\Large{\bf Response to Comment of Shukla and Akbari-Moghanjoughi }

\end{center}

\begin{center}
J. Vranjes$^1$, B. P. Pandey$^2$, and S. Poedts$^3$

$^{1}$Von Karman Institute, Waterloosesteenweg 72, 1640 Sint-Genesius-Rode, Brussels,
 Belgium.\\

 $^{2}$Department of Physics and Astronomy, Macquarie University, Sydney, NSW
2109, Australia.\\
$^{3}$Center for Plasma Astrophysics,  and Leuven Mathematical Modeling and Computational Science Center
(LMCC), K. U. Leuven, Celestijnenlaan 200B, 3001 Leuven,  Belgium.

E-mail: jvranjes@yahoo.com; drbp.pandey@gmail.com; Stefaan.Poedts@wis.kuleuven.be

\end{center}

\vspace{2cm}

{\bf Abstract:}

Shukla and Akbari-Moghanjoughi have {\it  corrected} their Comment (see their version 1 on `arXiv:1207.7029v1) to EPL on our work \cite{v1} after receiving our Response from the Editors of EPL. We have a pleasant duty  at hand to present our second Response to their second version of the Comment. It is hoped that this response adds strength to our plea {\it for a common sense} \cite{v1} on quantum description of plasmas.

\vspace{1cm}


\vspace{2cm}

Shukla and Akbari-Moghanjoughi have  uploaded their Comment to ArXiv before having an opinion of adjudicator from EPL. Then after reading our formal Response which they obtained from the Editors of EPL, they realized their objections were unjustified, so that they modified their Comment and submitted it to both EPL and ArXiv again. This forced us to act accordingly and this is our Response to their second version of the Comment.

We do not see the point of invoking some classical works on quantum phenomena as Shukla and Akbari-Moghanjoughi do in their Comment. The objective
of our Note \cite{v1} is clearly defined in our text: works on quantum plasma published in the past few years. Hence, we did not deal with works of {\em
``many
distinguished physicists"} who {\em  ``laid down foundation to collective
interactions in dense quantum plasmas about sixty years ago"}, and we also did not analyze {\em ``Nobel Prize wining papers''} from 80 years ago. Our concern is how the quantum theory is being used in plasma nowadays and  in particular within the fluid plasma theory.

We discussed  several  criteria in \cite{v1}  which should be checked before
using quantum effects in plasma physics.  These include the following: i) applicability of fluid theory (discussed in the first paragraph and in the table 1),  ii)
temperature-density relation which follows from  de Broglie-Heisenberg relations, iii)
the condition of weakly non-ideal plasma (discussed on page 2 and in table 1), and iv) electron degeneracy.

The items i) and iii) are essential  for the application of fluid equations, and they are discussed in relation with quantum effects in our work \cite{v1}.
However,  these items  have not been tackled at all by Shukla and Akbari-Moghanjoughi. Being unable to justify their objections and after reading our first Response, in their {\em second version} of the Comment Shukla and Akbari-Moghanjoughi write about
``metallic conductors'' and ``solid density plasma''. We observe that the term ``solid density'' can hardly have any meaning. On the other hand, in \cite{v1} we did not mentioned at all metallic conductors. As far as we know, Shukla and Akbari-Moghanjoughi themselves are not involved in research of  ``metallic conductors'', on the contrary they are well involved in so called fluid quantum plasma theory. So instead of concentrating on real issues, Shukla and Akbari-Moghanjoughi speak about metallic conductors which are out of scope.

Practically all the Comment by Shukla and Akbari-Moghanjoughi is devoted to the degeneracy temperature [which is our item iv) for which we used 8 lines only].

The worthlessness of objections by Shukla and Akbari-Moghanjoughi  may be seen from {\it their} criterion for the applicability of classical theory at room temperature:
\be
n\ll 10^{24} \,\,\mbox{m$^{-3}$}.
\label{e1}
\ee
  We note that, for example, the number density for air at room temperature is around $3\cdot 10^{25}$  m$^{-3}$ which is greater than the value given above. Yet, we do not recall anybody using quantum theory to discuss dynamics of air and terrestrial atmosphere.
 As Shukla and Akbari-Moghanjoughi claim with the number given in Eq.~(\ref{e1}), the air density and most of the world around requires quantum theory. In fact the true value that should be in Eq.~(\ref{e1}) is  around 4 orders of magnitude greater, i.e., $8.2 \cdot 10^{27}$ m$^{-3}$ as follows from our work \cite{v1},  and from the same  de Broglie-Heisenberg relations  used also by Shukla and Akbari-Moghanjoughi; see also the original `v1' version of their Comment (which is ever changing!!) submitted to ArXiv \cite{s}.

According to Shukla and Akbari-Moghanjoughi (see the first version of their Comment `v1') \cite{s}, degeneracy appears at temperature below
\be
T_d(K)=2.7\cdot 10^{-7} n^{2/3},
\label{e2}
\ee
 where $n$ is in $m^{-3}$. Taking for example a typical fusion plasma with
$n=10^{20}$ $m^{-3}$ yields  an enormously high threshold for degeneracy, $T_d=5.8 \cdot 10^6$ K. So according to Shukla and Akbari-Moghanjoughi every fusion
plasma or laboratory plasmas with  similar densities are completely quantum systems.
Clearly,   Eq.~(\ref{e2})  makes no physical sense. The correct value for the degeneracy temperature threshold follows from Fermi-Dirac statistics  that can be
found in any good book,  $T<[h^2/(2 m_e \kappa)] [n/(2.76 \pi)]^{2/3}$, and which consequently yields
\be
T< 4\cdot 10^{-15} n^{2/3}. \label{e4}
\ee
Here the temperature is in K and the number density in $m^{-3}$. For the same density as above $n=10^{20}$ $m^{-3}$ this yields $T<0.09$ K. Hence, in view of
such a low temperature threshold needed for quantum effects to appear it is clear that their importance is very limited, as we claimed in \cite{v1}; the typographical error $n^{3/2}$ instead of $n^{2/3}$, which Shukla and Akbari-Moghanjoughi refer to, changes nothing in physics and in our conclusions.

Due to unknown reasons Shukla and Akbari-Moghanjoughi invoke also the Saha's
ionization criterion into discussion and claim  that it ``dictates that in order for the plasma to form one must have sufficiently high degree of ionization." This is an absurd statement, and just for the benefit of Shukla and Akbari-Moghanjoughi we stress  that an  ionized gas is  cold  `plasma' if: 1) the Debye length is well below the system scale, 2)  there is enough number of particles per unit volume, and 3) the plasma frequency is well above the collision frequency. These conditions may be discussed without invoking neutrals, therefore  the Saha formula has nothing to do with this. The Saha's
ionization formula merely determines the ionization ratio in a mixture containing gas and charged species, and this only on condition of thermodynamic equilibrium and by assuming an ideal gas, i.e., containing non-interacting species so that the total partition function is just the product of partition functions of all species. In other words, Coulomb interaction between the species (which implies the collective behavior) is completely ignored. For that reason the Saha equation cannot possibly `dictate' the forming of a plasma, although  it nicely describes some of its properties for very weakly ionized gases. This issue however is completely out of scope of our work \cite{v1}, and Shukla and Akbari-Moghanjoughi would do better without it. This is because they talk about ridiculously low ionization of air at room temperature and yet in the same breath they immediately contradict themselves by writing  about low temperature plasmas at room temperature, from which they derive their wrong criterion  given in Eq.~(\ref{e1}) above.

 In their lengthy discussion {\em related to degeneracy} Shukla and Akbari-Moghanjoughi `conclude' that our `figure and the table displayed in Ref. \cite{v1}
are fallacious, and are of no use for defining the regimes
of quantum plasmas.' We observe that our Figure and Table had nothing to do with degeneracy [the criterion iv)] but with criteria i), ii) and iii) mentioned above, and both Figure and Table are completely correct and in agreement with any plasma physics textbook. Shukla and Akbari-Moghanjoughi did not provide any argument that would prove the opposite. On the contrary,  Eq.~(\ref{e1}) given above contains a trivially wrong condition given by  Shukla and Akbari-Moghanjoughi so it cannot be used against our results.

To make this Response self-contained and easy to read, here we provide the values of constants used above:  $\kappa=1.3807 \cdot10^{-23}$ J/K, $m_e=9.1094 \cdot
10^{-31}$ kg, $h=6.6261\cdot 10^{-34}$ Js, $\hbar=1.0546\cdot 10^{-34}$ Js.

We conclude that a) Shukla and Akbari-Moghanjoughi ignored our criteria i) and iii) which are essential for application of fluid dynamics with quantum effects,
and b) their criticism of our criteria ii)  and iv) is against elementary facts.


\end{document}